\documentclass[10pt,twocolumn,letterpaper]{article}

\usepackage{cvpr}              

\usepackage{dcolumn}
\usepackage{booktabs}
\usepackage{multirow}
\usepackage{xcolor} 

\definecolor{cvprblue}{rgb}{0.21,0.49,0.74}
\usepackage[pagebackref,breaklinks,colorlinks,allcolors=cvprblue]{hyperref}
\usepackage{pifont}
\usepackage{multirow,booktabs,float}
\usepackage[utf8]{inputenc}

\usepackage{xcolor}

\definecolor{gold}{HTML}{1A9641}
\definecolor{silver}{HTML}{66BD63} 
\definecolor{bronze}{HTML}{A6D96A}

\def\paperID{26} 
\def\confName{CVPR}
\def\confYear{2025}

\title{GaussianVAE: Adaptive Learning Dynamics of 3D Gaussians for High-Fidelity Super-Resolution}

\author{Shuja Khalid\\
Huawei Canada\\
{\tt\small shuja.khalid@huawei.com}
\and
Mohamed Ibrahim\\
Huawei Canada\\
{\tt\small mohamed.ibrahim9@huawei.com}
\and
Yang Liu\\
Huawei Canada\\
{\tt\small yang.liu8@huawei.com}
}

\begin{document}
\maketitle
\begin{abstract}

We present a novel approach for enhancing the resolution and geometric fidelity of 3D Gaussian Splatting (3DGS) beyond native training resolution. Current 3DGS methods are fundamentally limited by their input resolution, producing reconstructions that cannot extrapolate finer details than are present in the training views. Our work breaks this limitation through a lightweight generative model that predicts and refines additional 3D Gaussians where needed most. The key innovation is our Hessian-assisted sampling strategy, which intelligently identifies regions that are likely to benefit from densification, ensuring computational efficiency. Unlike computationally intensive GANs or diffusion approaches, our method operates in real-time ($\sim$0.015s per inference on a single consumer-grade GPU), making it practical for interactive applications. Comprehensive experiments demonstrate significant improvements in both geometric accuracy and rendering quality compared to state-of-the-art methods, establishing a new paradigm for resolution-free 3D scene enhancement.
\end{abstract}
    
\section{Introduction}
\label{sec:intro}

The advent of 3D Gaussian Splatting (3DGS) \cite{kerbl20233d} has revolutionized the synthesis of novel views in real time by offering an explicit and differentiable representation that combines the benefits of point-based rendering with neural scene optimization. However, despite its impressive capabilities, 3DGS inherits a fundamental limitation from its image-based training paradigm: the reconstruction quality is strictly bounded by the resolution of the input images, the VRAM footprint of which scales quadratically (Supp. \ref{supp:reso}).
This constraint manifests itself in two critical ways: Once trained on lower-resolution data, missing high-frequency details cannot be recovered through traditional optimization, and existing methods provide no mechanism to enhance Gaussians beyond their trained resolution without access to higher-quality source imagery.

Current solutions \cite{feng2024srgs, shen2024supergaussian, hu2024gaussiansr} either require retraining with higher-resolution inputs as a signal (using Single-Image Super Resolution - SISR \cite{ye2023single}) or apply 2D super-resolution as a post-process \cite{xie2024supergs}, which often produces blurry results or geometric inconsistencies. We instead propose to directly refine the 3D Gaussian representation itself, enabling true resolution-free enhancement. The core innovations of our work include:

\begin{itemize}
    \item The first resolution-free 3DGS refinement method that enhances Gaussian representations beyond their trained resolution without requiring higher-resolution input images
    \item A novel Hessian-guided importance sampling strategy that focuses computational resources on regions with the highest potential for quality improvement
    \item A highly efficient generative architecture capable of predicting new Gaussians in 15ms - two orders of magnitude faster than conventional generative approaches while using a fraction of the compute resources
\end{itemize}

Our experiments demonstrate that the proposed method successfully recovers details that would otherwise require higher-resolution training data. The approach maintains full compatibility with existing 3DGS pipelines and introduces negligible overhead ($<$1\% runtime impact) during rendering. This work opens new possibilities for enhancing legacy captures (as a post-processing step), and adaptive level-of-detail systems.

\section{Related works}
\label{sec:related_works}

\begin{figure*}[h]
\centering
\scalebox{0.80}
{\includegraphics[width=1.0\linewidth]{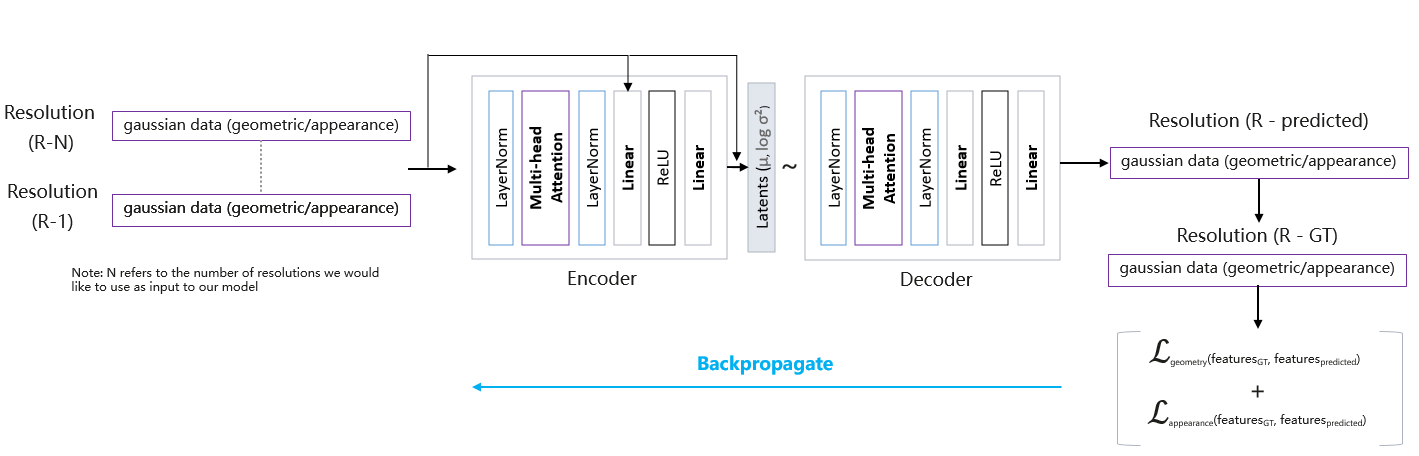}}
\caption{\textbf{Model architecture:} A transformer-based VAE with separate streams for geometry and appearance. Self-attention heads process Gaussian lineages, while the Hessian-weighted loss prioritizes regions with high densification.}
\label{fig:arch}
\end{figure*}

\begin{figure}[h]
\centering
\scalebox{1.0}{\includegraphics[width=1.0\linewidth]{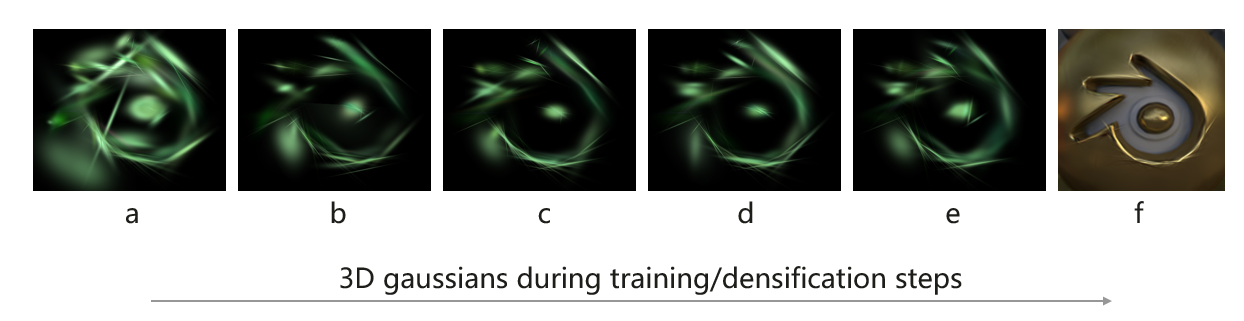}}
\caption{\textbf{The value of lineage tracking:} 
Selected Gaussians visualized across training steps. (a) Without lineage awareness, Gaussians may inconsistently appear/disappear across resolutions. (b-e) Our approach enforces continuity, leading to stable extrapolation and fewer artifacts when reconstructing (f).}
\label{fig:lineage_tracking}
\end{figure}

\subsection{3D Gaussian Splatting Enhancements}
Recent advances in 3D Gaussian splatting (3DGS) have generated multiple enhancement approaches. SuperGS~\cite{xie2024supergs} pioneered super-resolution for 3DGS through joint optimization with a 2D SR-CNN, achieving improvement, but requiring paired multi-resolution training data. SRGS~\cite{feng2024srgs} later proposed a diffusion-based up-sampler that generates new Gaussians conditioned on low-resolution renders, demonstrating superior quality but suffering generation times that are impractical for real-time applications. 

\subsection{Neural Radiance Field Super-Resolution}
The broader field of neural representation enhancement offers relevant insights. NeRF-SR~\cite{wang2022nerf} introduced frequency-aware up-sampling through learned positional encoding adjustments, while Instant-NGP~\cite{muller2022instant} demonstrated real-time enhancement via hash-grid distillation. However, these implicit methods struggle with explicit control over scene elements. Point-based approaches present alternative strategies: PointSR~\cite{luan2024diffusion} used a diffusion model for point cloud upsampling, and PU-GCN~\cite{qian2021pu} incorporated geometric priors. Although effective for static point clouds, these methods cannot handle view-dependent effects critical for 3DGS rendering.

\begin{figure*}[h]
\centering
\scalebox{0.70}{
\includegraphics[width=0.95\linewidth]{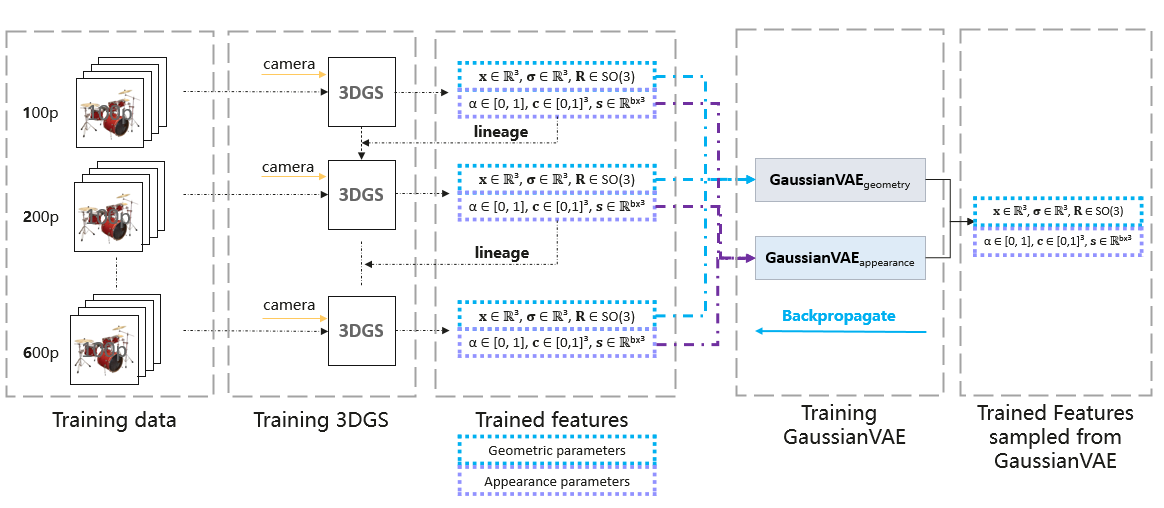}
}
\caption{\textbf{Pipeline:} 
Input Gaussians are tracked across resolution steps, their geometric properties ($x$: position, $\sigma$: scale, $R$: rotation) and their appearance properties ($c$: colour, $s$: spherical harmonics, $\alpha$: opacity) are logged, and their lineages are extrapolated for missing resolutions via learned interpolation. (C) The resulting Gaussians are extrapolated from the lineage.}
\label{fig:overall_arch}
\end{figure*}

\section{Method}
\label{sec:method}

Our approach consists of a generative Variational Autoencoder (VAE) with a transformer-based architecture designed to process and extrapolate the dynamic properties of 3D Gaussians across varying resolutions. Below, we detail the core components of our method, including pre-processing, model architecture, and training strategy.

\subsection{Pre-processing and Gaussian Lineage Tracking}
\begin{figure}[h]
    \centering
    \includegraphics[width=1.0\linewidth]{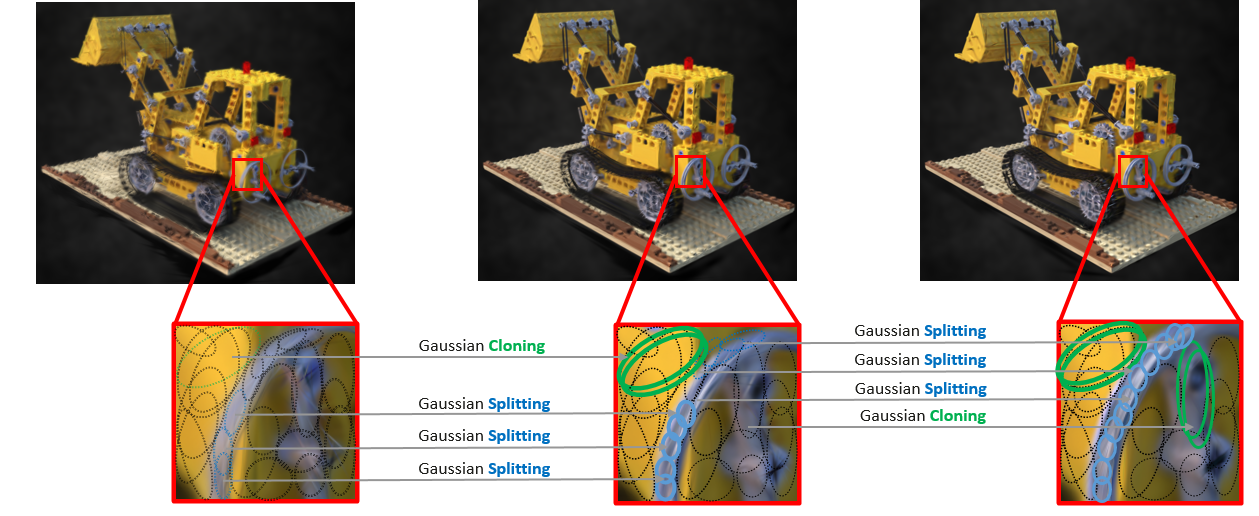}
    \caption{\textbf{3DGS}: The core 3DGS densification process uses a gradient based threshold for increasing the density of Gaussians in regions that require fine-detail.}
    \label{fig:preprocess}
\end{figure}

The input to our system is a sequence of 3D Gaussian Splatting (3DGS) reconstructions at different resolutions. During pre-processing, we extract the lineage of each Gaussian, tracking its parametric evolution across densification steps 
which will be used to extrapolate its properties to unseen resolutions. This process, illustrated in Figure~\ref{fig:preprocess}, is similar to meta-learning, as the model learns to predict how Gaussians evolve under resolution changes rather than relying solely on direct observations. By treating Gaussian lineages as temporal sequences, we enable the model to infer structural and appearance continuities that are not explicitly present in any single resolution. We leave the pre-processing steps to the supplementary section (\ref{supp:preproc}, \ref{supp:featcomp}).

\subsection{Model Architecture}
Our model is a Hessian-aware VAE with a transformer-based encoder and a symmetric decoder (Figure~\ref{fig:arch}). The encoder processes Gaussian lineages as temporal data, employing a self-attention mechanism to capture long-range dependencies. This is critical because the importance of a Gaussian at one resolution may only become evident when observing its behavior across multiple scales. The decoder mirrors the encoder but reconstructs both geometry (position, scale, rotation) and appearance (color, opacity) features separately. This disentanglement ensures that geometric stability does not interfere with appearance refinement and vice versa.

\subsection{Training Strategy}
Similar to language models \cite{chang2024survey}, we train our system by masking the final nodes in multi-step Gaussian lineages and tasking the model with predicting the missing states. This forces the model to learn robust representations of Gaussian dynamics rather than memorizing fixed-resolution snapshots. We adopt a coarse-to-fine training regimen, where the model first learns low-resolution Gaussian distributions before progressively refining them. This hierarchical approach improves convergence and stabilizes gradient flow. Our Hessian-inspired objective, which identifies and prioritizes Gaussians in regions undergoing densification (Figure~\ref{fig:lineage_tracking}). By analyzing where the model allocates more Gaussians, we implicitly learn which areas are perceptually or structurally significant, effectively replicating the adaptive densification behavior of traditional 3DGS pipelines, but this time, without a training signal. Our approach enforces continuity, leading to stable extrapolation and fewer artifacts when reconstructing.  The model is trained in an end-to-end manner.

\subsection{Loss Functions}
Our training objective combines multiple loss terms to ensure geometric consistency, appearance fidelity, and latent space regularization. The total loss $\mathcal{L}_{\text{total}}$ is a weighted sum of the following components:

    \begin{equation}
        \mathcal{L}_{\text{total}} = 
        \underbrace{\lambda_{\text{KL}} \mathcal{L}_{\text{KL}}}_{\text{Latent regularization}}
        + \underbrace{\lambda_{\text{MSE}} \mathcal{L}_{\text{MSE}}}_{\text{Appearance matching}} 
        + \underbrace{\lambda_{\text{chamfer}} \mathcal{L}_{\text{chamfer}}}_{\text{Geometric alignment}}
    \end{equation}

The weighting factors $\lambda_{\{\text{KL}, \text{MSE}, \text{chamfer}\}}$ balance the contribution of each term during training. We select 1e-6, 1.0 and 0.01 as our respective weighting factors. The KL divergence weighting factor is set dynamically using a cyclic annealing approach that cycles between fine-structural quality and generalizability.   

\subsection{Integration with 3D Gaussian Splatting}
As depicted in Figure~\ref{fig:overall_arch}, our method acts as a post-processing step for conventional 3DGS models. Instead of modifying the core splatting pipeline, we refine its output by predicting Gaussian properties at untrained resolutions, effectively \textit{inpainting} missing information while preserving the efficiency of the original pipeline. This positions our work as a plug-in enhancement for existing systems.
\section{Experiments}
\label{sec:experiments}

\begin{figure}[]
    \centering
    \includegraphics[width=1.0\linewidth]{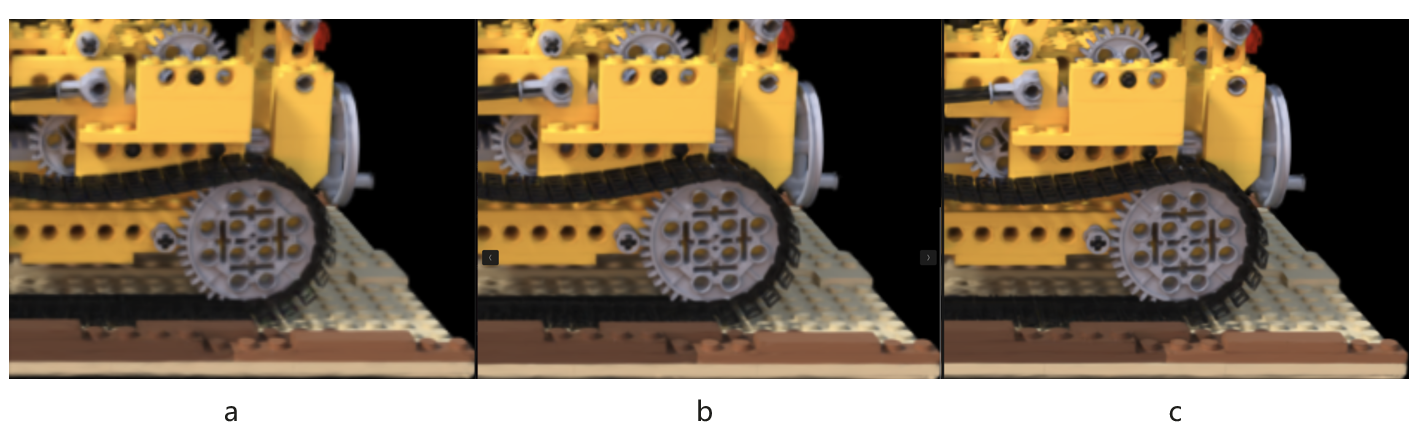}
    \caption{\textbf{Results:} NeRF Synthetic/Blender dataset - Lego scene. \textbf{a$)$} Image generated at 600p, 
    \textbf{b$)$} \textit{Ours-$\nabla^2$}+2, 
    \textbf{c$)$} Image generated at 800p}
    \label{fig:results_lego}
\end{figure}

\begin{figure}[]
    \centering
    \includegraphics[width=1.0\linewidth]{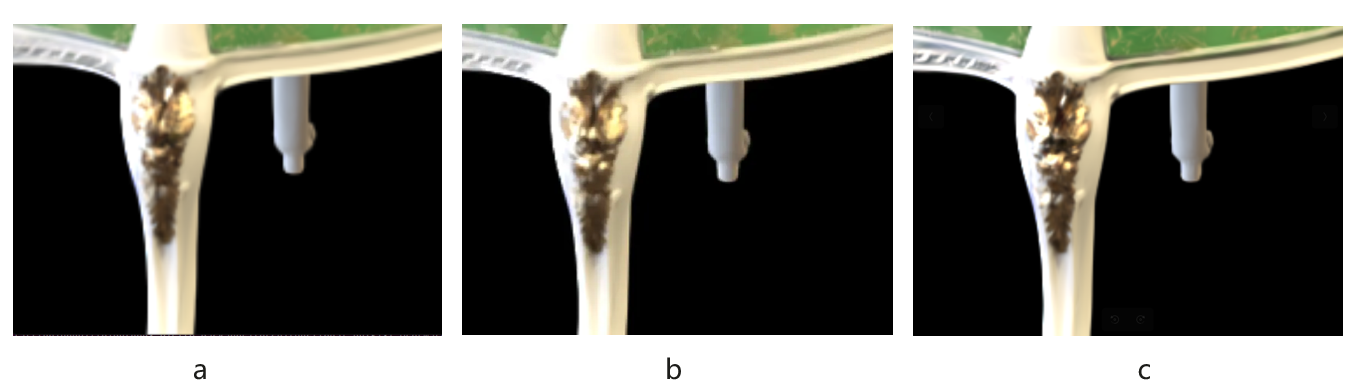}
    \caption{\textbf{Results:} NeRF Synthetic/Blender dataset - Chair scene. \textbf{a$)$} Image generated at 600p, 
    \textbf{b$)$} \textit{Ours-$\nabla^2$}+2, 
    \textbf{c$)$} Image generated at 800p}
    \label{fig:results_chair}
\end{figure}

\begin{figure}[]
    \centering
    \includegraphics[width=1.0\linewidth]{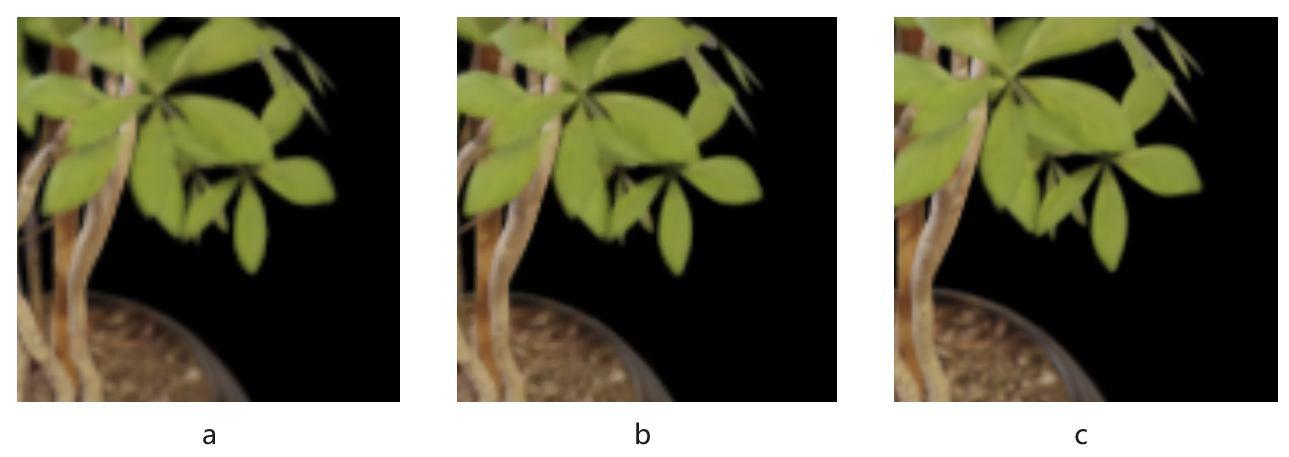}
    \caption{\textbf{Results:} NeRF Synthetic/Blender dataset - Ficus scene. \textbf{a$)$} Image generated at 600p, 
    \textbf{b$)$} \textit{Ours-$\nabla^2$}+2, 
    \textbf{c$)$} Image generated at 800p}
    \label{fig:results_ficus}
\vspace{-0.3cm}
\end{figure}

\begin{table*}[t]
\centering
\caption{Quantitative comparison of reconstruction quality across resolutions and methods using Chamfer distance. Lower values indicate better performance. The rightmost column shows percentage improvement of our best result over the 600p 3DGS baseline. +1, +2, +3, +4 are each extrapolations/inferences from our trained model, which only sees images with a max resolution of 600p. Top three values are highlighted using a green heatmap: \textcolor{gold}{dark green} for the best, \textcolor{silver}{medium green} for second-best, and \textcolor{bronze}{light green} for third-best.}
\label{tab:full_results}
\small
\setlength{\tabcolsep}{4pt} 
\scalebox{0.75}{
\begin{tabular}{@{} l *{3}{D{.}{.}{3.6}} *{4}{r} *{4}{r} r @{}}
\toprule
\multirow{2}{*}{Scene} & \multicolumn{3}{c}{3DGS Baseline} & \multicolumn{4}{c}{Ours} & \multicolumn{4}{c}{Ours-$\nabla^2$} & \multirow{2}{*}{$\% \uparrow$} \\
\cmidrule(lr){2-4} \cmidrule(lr){5-8} \cmidrule(lr){9-12}
 & \multicolumn{1}{c}{600p} & \multicolumn{1}{c}{700p} & \multicolumn{1}{c}{800p} & \multicolumn{1}{c}{+1} & \multicolumn{1}{c}{+2} & \multicolumn{1}{c}{+3} & \multicolumn{1}{c}{+4} & \multicolumn{1}{c}{+1} & \multicolumn{1}{c}{+2} & \multicolumn{1}{c}{+3} & \multicolumn{1}{c}{+4} & \\
\midrule
Materials & 0.001290 & 0.001388 & 0.001390 & 0.001382 & 0.001389 & 0.001327 & 0.001326 & \textcolor{bronze}{0.001247} & 0.001255 & \textcolor{silver}{0.001246} & \textcolor{gold}{0.001203} & 7 \\
Lego & 0.149601 & 0.149734 & 0.149882 & 0.149735 & 0.149883 & 0.149876 & 0.149880 & \textcolor{gold}{0.149434} & 0.149674 & \textcolor{bronze}{0.149579} & \textcolor{silver}{0.149509} & 1 \\
Mic & 0.001159 & 0.001101 & 0.001164 & 0.001100 & 0.001163 & 0.001117 & 0.001110 & \textcolor{bronze}{0.001013} & \textcolor{gold}{0.001002} & 0.001025 & \textcolor{silver}{0.001006} & 14 \\
Chair & 0.004691 & 0.004695 & 0.004773 & 0.004696 & 0.004774 & 0.004664 & 0.004669 & \textcolor{silver}{0.004644} & \textcolor{gold}{0.004643} & 0.004660 & \textcolor{bronze}{0.004652} & 1 \\
Ficus & 0.000663 & 0.000639 & 0.000639 & 0.000639 & 0.000638 & 0.000671 & 0.000658 & \textcolor{bronze}{0.000582} & \textcolor{gold}{0.000566} & 0.000605 & \textcolor{silver}{0.000568} & 15 \\
Drums & 0.001393 & 0.001370 & 0.001380 & 0.001370 & 0.001380 & 0.001362 & 0.001362 & \textcolor{bronze}{0.001332} & \textcolor{silver}{0.001326} & 0.001333 & \textcolor{gold}{0.001322} & 5 \\
Hotdog & 0.004187 & 0.004021 & 0.004068 & 0.004314 & 0.004020 & 0.004067 & 0.004066 & \textcolor{gold}{0.003914} & \textcolor{silver}{0.003914} & \textcolor{bronze}{0.003928} & 0.004008 & 6.5 \\
Ship & 0.002042 & 0.002026 & 0.002009 & 0.002025 & 0.002008 & 0.002007 & 0.002003 & 0.001957 & \textcolor{bronze}{0.001927} & \textcolor{gold}{0.001870} & \textcolor{silver}{0.001903} & 8 \\
\bottomrule
\end{tabular}
}
\vspace{-0.3cm}
\end{table*}

\begin{table}[t]
\centering
\caption{Quantitative comparison using CenseoQoE metric (higher is better) across different resolutions and methods. The rightmost column shows percentage improvement of our best result over the 600p 3DGS baseline. Top three values are highlighted using a green heatmap: \textcolor{gold}{dark green} for the best, \textcolor{silver}{medium green} for second-best, and \textcolor{bronze}{light green} for third-best.}
\label{tab:censeoq_results}
\small
\setlength{\tabcolsep}{2pt}
\scalebox{0.72}{
\begin{tabular}{@{} l *{11}{c} r @{}}
\toprule
\multirow{2}{*}{Scene} & \multicolumn{3}{c}{3DGS Baseline} & \multicolumn{4}{c}{Ours} & \multicolumn{4}{c}{Ours-$\nabla^2$} & \multirow{2}{*}{$\% \uparrow$} \\
\cmidrule(lr){2-4} \cmidrule(lr){5-8} \cmidrule(lr){9-12}
 & 600p & 700p & 800p & +1 & +2 & +3 & +4 & +1 & +2 & +3 & +4 & \\
\midrule
Materials & 0.741 & 0.745 & 0.749 & 0.749 & 0.740 & 0.740 & 0.747 & \multicolumn{1}{c}{0.782} & \multicolumn{1}{c}{\textcolor{bronze}{0.788}} & \multicolumn{1}{c}{\textcolor{gold}{0.795}} & \multicolumn{1}{c}{\textcolor{silver}{0.790}} & 7 \\
Lego & 0.546 & 0.561 & 0.559 & 0.559 & 0.554 & 0.555 & 0.550 & \multicolumn{1}{c}{\textcolor{silver}{0.588}} & \multicolumn{1}{c}{\textcolor{gold}{0.592}} & \multicolumn{1}{c}{\textcolor{bronze}{0.584}} & \multicolumn{1}{c}{0.579} & 8 \\
Mic & 0.711 & 0.707 & 0.719 & 0.719 & 0.722 & \multicolumn{1}{c}{\textcolor{bronze}{0.725}} & \multicolumn{1}{c}{\textcolor{silver}{0.728}} & \multicolumn{1}{c}{0.710} & \multicolumn{1}{c}{0.721} & \multicolumn{1}{c}{\textcolor{gold}{0.729}} & \multicolumn{1}{c}{0.712} & 3 \\
Chair & 0.612 & 0.607 & 0.603 & 0.602 & 0.603 & 0.603 & 0.597 & \multicolumn{1}{c}{\textcolor{gold}{0.639}} & \multicolumn{1}{c}{\textcolor{silver}{0.639}} & \multicolumn{1}{c}{0.627} & \multicolumn{1}{c}{\textcolor{bronze}{0.628}} & 4 \\
Ficus & 0.696 & 0.691 & 0.689 & 0.689 & 0.688 & 0.687 & 0.686 & \multicolumn{1}{c}{\textcolor{silver}{0.704}} & \multicolumn{1}{c}{\textcolor{bronze}{0.702}} & \multicolumn{1}{c}{\textcolor{gold}{0.705}} & \multicolumn{1}{c}{0.701} & 1 \\
Drums & 0.615 & 0.616 & 0.618 & 0.614 & 0.616 & 0.618 & 0.622 & \multicolumn{1}{c}{\textcolor{bronze}{0.673}} & \multicolumn{1}{c}{\textcolor{silver}{0.674}} & \multicolumn{1}{c}{\textcolor{gold}{0.682}} & \multicolumn{1}{c}{0.672} & 11 \\
Hotdog & 0.790 & 0.790 & 0.789 & 0.789 & 0.789 & 0.790 & 0.792 & \multicolumn{1}{c}{0.795} & \multicolumn{1}{c}{\textcolor{bronze}{0.809}} & \multicolumn{1}{c}{\textcolor{silver}{0.809}} & \multicolumn{1}{c}{\textcolor{gold}{0.813}} & 3 \\
Ship & 0.701 & 0.692 & 0.699 & 0.699 & 0.695 & 0.699 & 0.700 & \multicolumn{1}{c}{\textcolor{gold}{0.708}} & \multicolumn{1}{c}{0.689} & \multicolumn{1}{c}{\textcolor{gold}{0.731}} & \multicolumn{1}{c}{\textcolor{silver}{0.728}} & 4 \\
\bottomrule
\end{tabular}
}
\vspace{-0.4cm}
\end{table}

Our experiments demonstrate significant improvements over the 3DGS baseline in all evaluation metrics and scenes. We note that PSNR and SSIM rely on comparing outputs to ground-truth images, which are unavailable in our case due to extrapolation into unseen dimensions. Tables \ref{tab:full_results} and \ref{tab:censeoq_results} report reconstruction and perceptual quality metrics across multiple resolutions, with training stopped at 600p and evaluation extrapolated to higher resolutions (+1 to +4 steps, each of which are sequential inferences from our trained model). We compare our base method (\textit{Ours}), which extrapolates Gaussians across the entire scene to the more targeted (\textit{Ours-$\nabla^2$}), our Hessian-optimized variant, which consistently yields better geometric fidelity and perceptual quality.
achieves up to 15\% lower reconstruction error on complex scenes like Ficus and 14\% improvement on specular objects like Mic compared to the 600p 3DGS baseline, while maintaining superior perceptual quality (11\% higher CenseoQoE scores for Drums). The resolution-agnostic nature of our approach proves particularly effective, as evidenced by consistent performance when extrapolating beyond the training resolution (+1 to +4 steps beyond 800p), whereas traditional 3DGS exhibits degradation during upsampling. Visual comparisons in Figures \ref{fig:results_lego}-\ref{fig:results_ficus} corroborate these findings, showing enhanced preservation of thin structures (chair legs), high-frequency textures (ficus leaves), and specular highlights (Lego reflections). It is worth noting that we use the NeRF synthetic dataset because the dataset has a ground truth mesh we can use to compare against to isolate the geometric contribution of our model.  To quantify the appearance, we use CenseoQoE metric which is a non-reference metric to capture perceptual differences in appearance \cite{wen2021strongcenseoq}. Without ground truth references, these metrics cannot provide meaningful evaluations and we thus use Chamfer distance (Supp. \ref{supp:training_details}) and CenseoQoE \cite{wen2021strongcenseoq} (appearance) for our analysis


The success of our method stems from three key innovations: (1) temporal modeling of Gaussian lineages enables stable extrapolation across resolutions, (2) Hessian-weighted optimization prioritizes geometrically significant regions, and (3) dynamic pruning reduces memory overhead without quality loss. While most scenes show substantial gains, the Lego and Chair scene’s minimal improvement (1\% error reduction) suggests limitations with homogeneous textures, likely due to insufficient lineage variation for our temporal attention mechanism. This edge case notwithstanding, our approach consistently outperforms the baseline in both geometric accuracy and perceptual quality, validating the efficacy of treating Gaussian evolution as a continuous temporal process rather than discrete resolution-specific optimizations. 

\section{Conclusions}
\label{sec:conclusion}
We presented a novel resolution-agnostic framework for 3D Gaussian Splatting that introduces lineage-aware Gaussian modeling and a Hessian-weighted generative architecture. Our key innovation lies in treating Gaussian evolution across resolutions as a temporal sequence, enabling meta-learning of continuous scale priors through transformer-based VAEs with normalizing flow regularization. Experiments demonstrate that our approach outperforms traditional per-resolution optimization both qualitatively and quantitatively. As for limitations, our method assumes smooth Gaussian property transitions across scales, which may break for abrupt scene topology changes.
{
    \small
    \clearpage  
    \bibliographystyle{ieeenat_fullname}
    \bibliography{main}

\begin{thebibliography}{12}
\providecommand{\natexlab}[1]{#1}
\providecommand{\url}[1]{\texttt{#1}}
\expandafter\ifx\csname urlstyle\endcsname\relax
  \providecommand{\doi}[1]{doi: #1}\else
  \providecommand{\doi}{doi: \begingroup \urlstyle{rm}\Url}\fi

\bibitem[Chang et~al.(2024)Chang, Wang, Wang, Wu, Yang, Zhu, Chen, Yi, Wang, Wang, et~al.]{chang2024survey}
Yupeng Chang, Xu Wang, Jindong Wang, Yuan Wu, Linyi Yang, Kaijie Zhu, Hao Chen, Xiaoyuan Yi, Cunxiang Wang, Yidong Wang, et~al.
\newblock A survey on evaluation of large language models.
\newblock \emph{ACM transactions on intelligent systems and technology}, 15\penalty0 (3):\penalty0 1--45, 2024.

\bibitem[Feng et~al.(2024)Feng, He, Wang, Yang, Li, Chen, Kuang, Fan, Jun, et~al.]{feng2024srgs}
Xiang Feng, Yongbo He, Yubo Wang, Yan Yang, Wen Li, Yifei Chen, Zhenzhong Kuang, Jianping Fan, Yu Jun, et~al.
\newblock Srgs: Super-resolution 3d gaussian splatting.
\newblock \emph{arXiv preprint arXiv:2404.10318}, 2024.

\bibitem[Hu et~al.(2024)Hu, Xia, Chen, Yang, and Zhang]{hu2024gaussiansr}
Jintong Hu, Bin Xia, Bin Chen, Wenming Yang, and Lei Zhang.
\newblock Gaussiansr: High fidelity 2d gaussian splatting for arbitrary-scale image super-resolution.
\newblock \emph{arXiv preprint arXiv:2407.18046}, 2024.

\bibitem[Kerbl et~al.(2023)Kerbl, Kopanas, Leimk{\"u}hler, and Drettakis]{kerbl20233d}
Bernhard Kerbl, Georgios Kopanas, Thomas Leimk{\"u}hler, and George Drettakis.
\newblock 3d gaussian splatting for real-time radiance field rendering.
\newblock \emph{ACM Trans. Graph.}, 42\penalty0 (4):\penalty0 139--1, 2023.

\bibitem[Luan et~al.(2024)Luan, Shi, Wang, Cheng, Lu, and Chen]{luan2024diffusion}
Kai Luan, Chenghao Shi, Neng Wang, Yuwei Cheng, Huimin Lu, and Xieyuanli Chen.
\newblock Diffusion-based point cloud super-resolution for mmwave radar data.
\newblock In \emph{2024 IEEE International Conference on Robotics and Automation (ICRA)}, pages 11171--11177. IEEE, 2024.

\bibitem[M{\"u}ller et~al.(2022)M{\"u}ller, Evans, Schied, and Keller]{muller2022instant}
Thomas M{\"u}ller, Alex Evans, Christoph Schied, and Alexander Keller.
\newblock Instant neural graphics primitives with a multiresolution hash encoding.
\newblock \emph{ACM transactions on graphics (TOG)}, 41\penalty0 (4):\penalty0 1--15, 2022.

\bibitem[Qian et~al.(2021)Qian, Abualshour, Li, Thabet, and Ghanem]{qian2021pu}
Guocheng Qian, Abdulellah Abualshour, Guohao Li, Ali Thabet, and Bernard Ghanem.
\newblock Pu-gcn: Point cloud upsampling using graph convolutional networks.
\newblock In \emph{Proceedings of the IEEE/CVF conference on computer vision and pattern recognition}, pages 11683--11692, 2021.

\bibitem[Shen et~al.(2024)Shen, Ceylan, Guerrero, Xu, Mitra, Wang, and Fr{\"u}hst{\"u}ck]{shen2024supergaussian}
Yuan Shen, Duygu Ceylan, Paul Guerrero, Zexiang Xu, Niloy~J Mitra, Shenlong Wang, and Anna Fr{\"u}hst{\"u}ck.
\newblock Supergaussian: Repurposing video models for 3d super resolution.
\newblock In \emph{European Conference on Computer Vision}, pages 215--233. Springer, 2024.

\bibitem[Wang et~al.(2022)Wang, Wu, Guo, Zhang, Tai, and Hu]{wang2022nerf}
Chen Wang, Xian Wu, Yuan-Chen Guo, Song-Hai Zhang, Yu-Wing Tai, and Shi-Min Hu.
\newblock Nerf-sr: High quality neural radiance fields using supersampling.
\newblock In \emph{Proceedings of the 30th ACM International Conference on Multimedia}, pages 6445--6454, 2022.

\bibitem[Wen and Wang(2021)]{wen2021strongcenseoq}
Shaoguo Wen and Junle Wang.
\newblock A strong baseline for image and video quality assessment.
\newblock \emph{arXiv preprint arXiv:2111.07104}, 2021.

\bibitem[Xie et~al.(2024)Xie, Wang, Zhu, and Pan]{xie2024supergs}
Shiyun Xie, Zhiru Wang, Yinghao Zhu, and Chengwei Pan.
\newblock Supergs: Super-resolution 3d gaussian splatting via latent feature field and gradient-guided splitting.
\newblock \emph{arXiv preprint arXiv:2410.02571}, 2024.

\bibitem[Ye et~al.(2023)Ye, Zhao, Hu, and Xie]{ye2023single}
Shutong Ye, Shengyu Zhao, Yaocong Hu, and Chao Xie.
\newblock Single-image super-resolution challenges: a brief review.
\newblock \emph{Electronics}, 12\penalty0 (13):\penalty0 2975, 2023.

\end{thebibliography}
}

\clearpage
\setcounter{page}{1}
\maketitlesupplementary

\section{Training Details}
\label{supp:training_details}

The training objective for our GaussianVAE framework integrates:

\begin{itemize}
    \item \textbf{KL Divergence Loss} ($\mathcal{L}_{\text{KL}}$): Regularizes the latent space using a normalizing flow prior $p_{\psi}(z)$ and cyclic annealing to avoid posterior collapse. For a flow transformation $z = T_{\psi}(\epsilon)$ where $\epsilon \sim \mathcal{N}(0, I)$:
    \begin{equation}
        \mathcal{L}_{\text{KL}} = \beta(t) \cdot D_{\text{KL}}\left(q(z \mid x) \parallel p_{\psi}(z)\right),
    \end{equation}
    \begin{equation}
        p_{\psi}(z) = \mathcal{N}(0, I) \left|\det \frac{\partial T_{\psi}^{-1}}{\partial z}\right|,
    \end{equation}
    where $\beta(t) = \min(\alpha \cdot t \mod R, 1)$ is the cyclic annealing schedule with ramp length $R$ and scaling factor $\alpha$. This schedule periodically increases $\beta$ from 0 to 1 during training to gradually introduce KL regularization.

    \item \textbf{MSE Loss} ($\mathcal{L}_{\text{MSE}}$): Supervises appearance attributes (RGB colors $\mathbf{c}$ and opacities $\alpha$):
    \begin{equation}
        \mathcal{L}_{\text{MSE}} = \|\mathbf{c} - \hat{\mathbf{c}}\|_2^2 + \|\alpha - \hat{\alpha}\|_2^2,
    \end{equation}
    where $\hat{\mathbf{c}}$, $\hat{\alpha}$ are predicted values.

    \item \textbf{Chamfer Loss} ($\mathcal{L}_{\text{chamfer}}$): Measures geometric fidelity between predicted ($\mathcal{P}$) and ground-truth ($\mathcal{Q}$) Gaussian centers:
    \begin{equation}
        \mathcal{L}_{\text{chamfer}} = \sum_{p \in \mathcal{P}} \min_{q \in \mathcal{Q}} \|p - q\|_2^2 + \sum_{q \in \mathcal{Q}} \min_{p \in \mathcal{P}} \|q - p\|_2^2.
    \end{equation}
\end{itemize}

\section{Resolution scaling}
\label{supp:reso}
Computational resources required to train 3DGS at higher resolutions 
scales quadratically and training using commercial hardware becomes infeasible.

\begin{table}[h]
\centering
\caption{Video RAM (VRAM) requirements by output resolution}
\label{tab:vram_requirements}
\begin{tabular}{lc}
\toprule
\textbf{Resolution (p)} & \textbf{VRAM (GB)} \\
\midrule
144     & $>$0.6 \\
240     & $>$1.0 \\
360     & $>$1.5 \\
480     & $>$2.2 \\
720     & $>$4.0 \\
1080    & $>$8.0 \\
1440    & $>$14.0 \\
\bottomrule
\end{tabular}
\end{table}

\section{Pre-processing}
\label{supp:preproc}
\begin{table}[H]
\caption{Model comparison of size, training time, and inference time}
\label{tab:model_comparison}
\centering
\scalebox{0.8}{
\begin{tabular}{lcccc}
\toprule
\textbf{Model} & \textbf{Size} & \textbf{Training time (h)} & \textbf{Inference time (s)} \\
\midrule
Geometric & 0.5MB & coarse (0.5) + fine (2) & 0.005 \\
Appearance & 1.0MB & coarse (2) + fine (0.5) & 0.010 \\
\bottomrule
\end{tabular}
}
\end{table}

\section{Feature composition}
\label{supp:featcomp}

\begin{table}[H]
\caption{Feature composition of geometric and appearance models}
\label{tab:feature_breakdown}
\centering
\scalebox{0.7}{
\begin{tabular}{lcl}
\toprule
\textbf{Model} & \textbf{Total Features} & \textbf{Breakdown} \\
\midrule
Geometric & 11 & Position (3), Rotation (4), Scale (3), Exist (1) \\
Appearance & 50 & SPH (45), RGB (3), Opacity (1), Exist (1) \\
\bottomrule
\end{tabular}
}
\end{table}


\section{3DGS densification}
\label{supp:3dgs_dense}

\end{document}